\documentclass[a4paper]{article}
\usepackage{authblk}
\usepackage[dvipdfmx]{graphicx} 
\usepackage[top=30truemm,bottom=25truemm,left=25truemm,right=25truemm]{geometry}
\usepackage{dcolumn}
\usepackage{bm}
\usepackage{amsmath}
\usepackage{braket}
\usepackage{amsfonts}
\usepackage{amssymb}
\usepackage{color}
\usepackage{hyperref}
\hypersetup{colorlinks=true, citecolor= blue, linkcolor= blue, filecolor= blue, urlcolor= blue}
\usepackage{comment}
\usepackage{physics}
\usepackage {nonfloat}
\usepackage{multirow}
\usepackage[dvipsnames]{xcolor}
\usepackage{lineno}
\usepackage[numbers, sort]{natbib}
\usepackage[normalem]{ulem}
\usepackage{makecell}
\usepackage{mathrsfs}

\title{Impurity quadrupole moments as local probes \\ of flux sectors in the Kitaev spin liquid}
\author[1]{Masahiro O. Takahashi}
\affil[1]{Computational Materials Science Research Team, RIKEN Center for Computational Science (R-CCS), Hyogo 650-0047, Japan}
\author[2]{Wen-Han Kao}
\affil[2]{Department of Physics, University of Wisconsin-Madison, Madison, Wisconsin 53706, USA}
\author[3]{Satoshi Fujimoto}
\affil[3]{Department of Materials Engineering Science, The University of Osaka, Toyonaka 560-8531, Japan}
\author[4]{Natalia B. Perkins\footnote{nperkins@umn.edu}}
\affil[4]{School of Physics and Astronomy, University of Minnesota, Minneapolis, Minnesota 55455, USA}

\begin{document}

\maketitle

\begin{abstract}
Emergent fluxes play a central role in the low-energy properties of quantum spin liquids (QSLs), where they encode the underlying gauge structure and fractionalization of spins.
Here, we show that the quadrupole moment of magnetic impurities provides a direct probe of these flux configurations in QSLs and can be measured by local tunneling spectroscopy.
Employing the SO(6) Majorana representation for spin-3/2 impurity operators in the isotropic Kitaev spin liquid together with a self-consistent mean-field approximation for impurity-related terms, we show that the ground-state flux sector can be identified by discontinuous jumps of the impurity quadrupole moment at the flux sector transition points.
We also demonstrate that the quadrupole correlations between impurities under a magnetic field exhibit exponential decay, with decay rates that depend sensitively on the flux sector.
Furthermore, we discuss the stability of $\pi$-fluxes bound to impurities with respect to model parameters and internal flux configurations, and relate our findings to Lieb’s conjecture on flux configurations.
These results establish the quadrupole moments of magnetic impurities as a sensitive tool to study fractionalized excitations and flux physics in Kitaev magnets.
\end{abstract}

\section*{Introduction}
Gauge fluxes play a central role in many quantum systems and can often be observed through their coupling to external fields, with their effects appearing as geometric phases in quantum mechanics \cite{Cohen2019}. 
For example, magnetic fluxes can be detected through electromagnetic responses such as the Aharonov-Bohm effect or quantized vortices in superconductors. 
In quantum spin liquids (QSLs), by contrast, the relevant gauge fluxes are emergent and do not couple directly to electric or magnetic fields, making them much more difficult to detect. These fluxes arise from the fractionalization of spin degrees of freedom into emergent quasiparticles coupled to a gauge field \cite{Balents2010, Savary2017, Broholm2020}. They distinguish different sectors of the spin liquid and play a central role in its low-energy properties.

A paradigmatic example is the Kitaev model on the honeycomb lattice \cite{Kitaev2006}, which provides an exactly solvable realization of a QSL without relying on mean-field approximations \cite{Wen1991}. 
The emergent gauge field is $Z_2$ valued, and each plaquette hosts a discrete flux $w_p=\pm1$, corresponding to $0$ or $\pi$-flux through a plaquette of the lattice. 
 In the pristine model without any defects, the ground state is in the zero-flux sector where all plaquettes exhibit $w_p = +1$.
 Furthermore, the system becomes a non-Abelian spin liquid once the time-reversal symmetry (TRS) is broken, and the excited $\pi$-fluxes in this phase can bind localized Majorana zero modes (MZMs), reflecting the nontrivial topology of the fractionalized excitations.

The Kitaev model provides a particularly transparent framework for studying the interplay between defects and gauge fluxes in the ground state. 
This has motivated a wide range of studies of lattice defects, including exactly solvable cases such as vacancies, bond disorder, and amorphous models \cite{Willans2010, Willans2011, Santhosh2012, Udagawa2018, Kao2021vacancy, Kao2021localization, Nasu2021, Dantas2022, Takahashi2023, Kao2024, KaoPRL2024, Yatsuta2024, Xiao2025, Xiao_Takahashi2025, WeiyaoLi2026, Cassella2023, Grushin2023}, as well as situations involving higher-spin impurities \cite{Dhochak2010, Das2016, Vojta2016, Bauer2024, Takahashi2025}, mixed-spin lattices \cite{Koga2019, Natori2025}, and crystallographic defects \cite{Seth2025, Seth2025_SW, Borhani2025}.
Even when exact solvability is lost, these systems retain the essential ingredients of the Kitaev framework, namely the presence of local conserved quantities that protect the defining features of the spin-liquid state \cite{Baskaran2007}.

A central challenge is to identify the local flux configuration through experimentally accessible observables.
Flux sectors can be inferred from dynamical spin responses through flux-dependent spectral features \cite{Messio2010, Punk2014, Knolle2014, Knolle2015, Yoshitake2016, Gohlke2018, Ferrari2019, Hickey2019, Nomura2021, Wang2025, Poree2025}, but such signatures arise indirectly through the coupling between itinerant quasiparticles and the gauge background.
Defects and impurities offer another route because they can bind fluxes and develop local and static responses that also depend on the surrounding flux sector \cite{Kolezhuk2006, Willans2010, Willans2011, Das2016, Vojta2016, Udagawa2018, Chen2020, Nasu2021, He2022, Dantas2022, Takahashi2023, Yatsuta2024, Lu2024, Seth2025}.
For example, a vacancy in the Kitaev model binds a $\pi$-flux and induces low-energy states that strongly affect nearby magnetic responses \cite{Willans2010}.

In our previous work \cite{Takahashi2025}, we determined the ground-state flux sectors of spin-$S$ magnetic impurities in the Kitaev spin liquid. While the local dipole moment vanishes at zero magnetic field, a spin-$3/2$ impurity possesses an additional quadrupole degree of freedom, which can remain finite without breaking the local conserved quantities. Since binding a $Z_2$ flux modifies the surrounding Majorana hopping pattern and the impurity--spin-liquid hybridization, it is expected to reconstruct the local impurity state and hence its quadrupole moment. This motivates us to investigate whether the impurity quadrupole moment can serve as a local observable of the emergent flux degrees of freedom.

In the present work, we focus on spin-$3/2$ impurities using an SO(6) Majorana representation \cite{Natori2016} combined with a self-consistent mean-field treatment, which enables us to study larger systems than previous spin-based calculations. We find that the impurity quadrupole moment exhibits sharp discontinuities across transitions between the bound-flux and zero-flux sectors, providing a direct local signature of the flux sector. We further show that, under a magnetic field, the quadrupole correlations decay exponentially with a flux-sector-dependent decay length, and discuss the stability of impurity-bound $\pi$-fluxes in relation to Lieb’s conjecture \cite{Lieb1992, Lieb1993, Lieb1994}.

From the experimental perspective, the impurity quadrupole moment can, in principle, be probed by tunneling spectroscopy ~\cite{Carrega2020, Knolle2020, Konig2020, Udagawa2021, Bauer2023, Takahashi2023, KaoPRL2024, Kao2024, Bauer2024, Zhang2025_PRB, Zhang2025_npj, WeiyaoLi2026}. Specifically, we consider a spin-polarized scanning tunneling microscopy (STM) setup where the spin-polarized tunneling current can be magnetically coupled to the spin-3/2 impurity in the Kitaev spin liquid, and thus the tunneling conductance is related to the impurity quadrupole moment. 

\begin{figure}[t]
    \centering
    \includegraphics[width=0.5\linewidth]{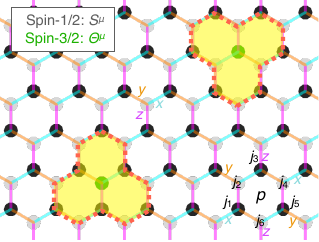}
    \caption{
    {\textbf{The Kitaev spin liquid with randomly distributed spin-3/2 impurities.}}
    Yellow plaquettes denote the presence of $\pi$-fluxes bound at impurity sites.
    Site labels are used for defining the flux operator on plaquette $p$.}
    \label{fig: fig1}
\end{figure}

\section*{Results}
\subsection*{Model}
\label{sec: model}
We start from the Kitaev model on the honeycomb lattice \cite{Kitaev2006}, in which spin-$1/2$ moments interact through bond-dependent Ising-type couplings.
Magnetic impurities are introduced by substituting a subset of spin-$1/2$ sites with spin-$3/2$ moments. We focus on the dilute limit where impurities are sufficiently separated such that the minimal loop surrounding an impurity site on the honeycomb lattice (a 12-site loop) does not contain any other impurities.
In this regime each impurity can be treated as an isolated defect with its own surrounding flux degrees of freedom.

The total Hamiltonian is written as
\begin{align}\label{Ham-total}
    \hat{H}_{\rm{total}} = \hat{H}_0 + \hat{H}_{\rm{SIA}} + \hat{H}_\Lambda + \hat{H}_\kappa,
\end{align}
where
\begin{align}
    \hat{H}_0 = -J\sum_{\begin{subarray}{c} j,k\notin \Lambda\\ \langle jk\rangle_\mu \end{subarray}}\hat{S}_j^\mu \hat{S}_k^\mu
\end{align}
represents the bulk Kitaev Hamiltonian acting on the spin-$1/2$ sites outside the impurity set $\Lambda$.
$\hat{S}_j^{\mu}$ denotes the $\mu$-component of the spin-1/2 at site $j$.
$\hat{H}_{\mathrm{SIA}}$ and $\hat{H}_\Lambda$ describe, respectively, the single-ion anisotropy (SIA) on impurity sites and the exchange coupling between impurity and its nearest-neighbor (NN) spins:
\begin{align}
    \hat{H}_{\rm{SIA}} = D_z\sum_{j\in\Lambda} (\hat{\Theta}_j^z)^2,\qquad
    \hat{H}_\Lambda = -g \sum_{\begin{subarray}{c} j\in\Lambda, k\notin\Lambda \\ \langle jk\rangle_\mu\end{subarray}} \hat{\Theta}_j^\mu \hat{S}_{k}^\mu.
\end{align}
Here, $\hat{\Theta}^\mu$ denotes the $\mu$-component of the spin-$3/2$ operator.
The SIA term naturally arises from the trigonal crystal-field distortion characteristic of spin-orbit-assisted $S = 3/2$  Mott insulators \cite{Peter2021} and has been investigated in both pure $S=3/2$ Kitaev spin liquids \cite{Jin2022, Natori2023} and staggered mixed-spin variants \cite{Natori2025}.
Here we include the SIA term only on impurity sites, with strength $D_z$.
We focus on positive $D_z$, which favors the $|m_z|=1/2$ doublet and, in the limit $D_z\to\infty$, provides a natural route to connect spin-3/2 impurities with an effective bond-disorder limit in the spin-1/2 Kitaev spin liquid.
All NN couplings we consider in this paper are Kitaev interactions, and their magnitudes, $J$ and $g$, are set to positive (ferromagnetic) values.
We set $J = 1$ as the reference energy scale.

We also consider the effect of an external magnetic field, assuming a uniform field $\bm{h}$ applied along the $[111]$ direction in the spin axis.
Within third-order perturbation theory, the field generates the effective interaction
\begin{align}
    \hat{H}_\kappa = \sum_{(jkl)}\kappa(\Delta_{\rm{flux}})~\hat{\mathcal{S}}_j^{\mu}\hat{\mathcal{S}}_k^\nu\hat{\mathcal{S}}_l^\rho,
    \label{eq: kappa-term}
\end{align}
where the operator $\hat{\mathcal{S}}_j^{\mu}$ 
represents either $\hat{S}_j^{\mu}$ or $\hat{\Theta}_j^\mu$ depending on the site $j$, and $(jkl)$ denotes triplets of sites that are mutually connected by NN bonds, and $(\mu,\nu,\rho)$ correspond to the bond directions $x,y,z$, respectively.
We note that the full Zeeman term generally breaks the exact conservation of the local flux operators and induces flux dynamics.
The coupling constant satisfies $\kappa(\Delta_{\rm{flux}})\propto{h^3}/{{(\Delta_{\rm{flux}})}^2}$,
where $h$ is the strength of the field and $\Delta_{\rm{flux}}$ denotes the excitation gap of conserved $Z_2$ fluxes.
In the bulk, we use $\Delta_{\rm{flux}}=0.065J$ \cite{Kitaev2006}.
However, when one of the sites in $(jkl)$ is an impurity site, we adopt a different value.
This value is estimated via a finite-size cluster analysis, as described in the section on finite-field results. 

The total Hamiltonian possesses a set of local conserved quantities, referred to as flux operators, each defined on a plaquette $p$.
Introducing the $\pi$-rotation operator about the $\alpha_j$-axis on a local spin site $j$, $\hat{R}_j^{\alpha_j}\equiv \exp[{i\pi \hat{\mathcal{S}}_j^{\alpha_j}}]$, the flux operator on plaquette $p$ is defined as
\begin{align}
    \hat{W}_p= - \prod_{j\in \partial p}\hat{R}_j^{\alpha_j}.
\end{align}
Here, $\partial p$ denotes the set of the six vertices surrounding plaquette $p$, and $\alpha_j$ specifies the direction of the outgoing bond from the plaquette at site $j$, as illustrated in Fig. \ref{fig: fig1}.
Since the flux operators commute both with the total Hamiltonian and mutually, $[\hat{W}_p,\, \hat{H}_{\rm{total}}]=[\hat{W}_p,\, \hat{W}_{p'}]=0$, their eigenvalues are conserved quantities taking values $\pm1$.

\begin{figure}[t]
    \centering
    \includegraphics[width=0.8\linewidth]{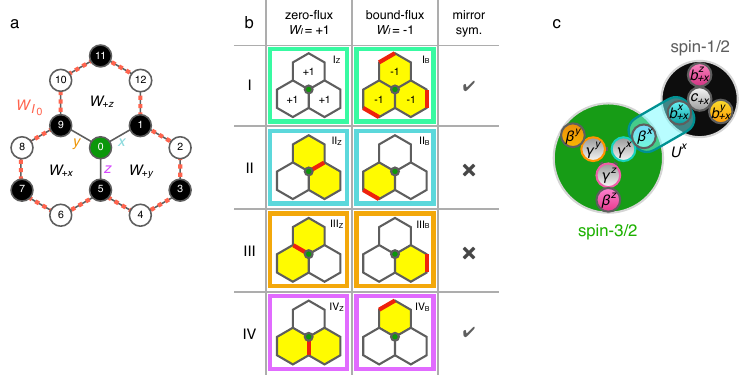}
    \caption{{\textbf{Internal flux operators.}}
    {\textbf{a}} Schematic illustration of the internal and triple-plaquette operators.
    Site labels correspond to those used in Eq. \eqref{eq: def_internal_W}.
    {\textbf{b}} Four patterns of internal flux configurations within a given flux sector.
    Yellow plaquettes denote internal $\pi$-fluxes.
    Thick red bonds indicate example choices of flipped gauge fields for each configuration; these choices are not unique in general.
    {\textbf{c}} Majorana representation for both spin-1/2 and spin-3/2 operators, showing the static gauge field on an $x$-bond as an example.}
    \label{fig: internal_flux}
\end{figure}

For spin-1/2 operators, the identity $\hat{R}_j^{\alpha_j}=-2iS_j^{\alpha_j}$ allows us to rewrite the flux operator $\hat{W}_p$ in the more familiar form:
\begin{align}
\hat{W}_p=2^6 \hat{S}_{j_1}^x \hat{S}_{j_2}^y \hat{S}_{j_3}^z \hat{S}_{j_4}^x \hat{S}_{j_5}^y \hat{S}_{j_6}^z.
\end{align}
For spin-3/2 operators, on the other hand, we adopt the pseudo-spin and pseudo-orbital representation \cite{Natori2016},
\begin{align}
    \hat{\Theta}^\mu = -\hat{\sigma}^\mu\otimes\left[\frac12 + \hat{T}^{\nu\rho}\right],
\end{align}
where $(\mu,\nu,\rho)$ is a cyclic permutation of $(x,y,z)$.
The pseudo-spin operators satisfy
\begin{align}
    \hat{\sigma}^\mu \equiv -i\exp[i\pi\hat{\Theta}^\mu],\quad [\hat{\sigma}^\mu, \hat{\sigma}^\nu]=2i\epsilon^{\mu\nu\rho}\hat{\sigma}^\rho.
\end{align}
The operator $\hat{T}^{\nu\rho}$ denotes the quadrupolar operator associated with the plane orthogonal to $\mu$. 
These operators are expressed in terms of three pseudo-orbital generators constructed from quadratic and cubic combinations of the spin-$3/2$ generators:
\begin{align}\label{Tdefin}
\begin{split}
    \hat{T}^x\equiv \frac{1}{\sqrt3}[(\hat{\Theta}^x)^2 - (\hat{\Theta}^y)^2], \quad \hat{T}^y\equiv \frac{2\sqrt3}{9}\,\overline{\hat{\Theta}^x\hat{\Theta}^y\hat{\Theta}^z},\quad
    \hat{T}^z=\hat{T}^{xy}\equiv (\hat{\Theta}^z)^2 - \frac54,\quad \hat{T}^{yz(zx)}\equiv-\frac{\hat{T}^z}{2} \pm\frac{\sqrt3 \hat{T}^x}{2},
\end{split}
\end{align}
in which the bar indicates a sum over all permutations
of the operators under it.
The pseudo-orbital operators satisfy $[\hat{T}^\mu,\,\hat{T}^\nu]=2i\epsilon^{\mu\nu\rho}\hat{T}^\rho$.
The pseudo-spin and pseudo-orbital sectors commute, $[\hat{\sigma}^\mu,\,\hat{T}^\nu]=0$.
This representation allows us to rewrite the flux operators that include the impurity site $0\in{\Lambda}$ as  the \textit{internal} flux operators:
\begin{align}
    \hat{W}_{+x}=2^5 \hat{S}_7^x \hat{S}_8^y \hat{S}_9^z \hat{\sigma}_0^x \hat{S}_5^y \hat{S}_6^z,\quad 
    \hat{W}_{+y}=2^5 \hat{S}_5^x \hat{\sigma}_0^y \hat{S}_1^z \hat{S}_2^x \hat{S}_3^y \hat{S}_4^z,\quad
    \hat{W}_{+z}=2^5 \hat{S}_9^x \hat{S}_{10}^y \hat{S}_{11}^z \hat{S}_{12}^x \hat{S}_1^y \hat{\sigma}_0^z,
    \label{eq: def_internal_W}
\end{align}
where $[\hat{W}_{+\mu}, \hat{H}_{\rm{total}}]=[\hat{W}_{+\mu}, \hat{W}_{+\nu}]=0$ with $\mu, \nu=x,y,z$, whose eigenvalues $\pm1$ are also conserved.
Note that the site labels used in Fig.~\ref{fig: internal_flux}a are impurity dependent and represent positions defined relative to the impurity site (0).

The product of the  internal flux operators yields
\begin{align}
    \hat{W}_{+x}\hat{W}_{+y}\hat{W}_{+z} = \hat{W}_{I_0} \equiv 2^{12}\prod_{k\in{\partial I_0}} \hat{S}_k^{\alpha_k}.
    \label{eq: Triple_plaquette}
\end{align}
Here $\partial I_j$ with $j=0$ denotes the set of twelve spin-$1/2$ sites surrounding a given impurity site $0\in\Lambda$, as labeled in Fig. \ref {fig: internal_flux}a.
The triple-plaquette operator $\hat{W}_{I_j}$ with $j\in\Lambda$ in general, whose eigenvalues are also conserved as $\pm1$ since $[\hat{W}_{I_j}, \hat{H}_{\rm{total}}]=0$, is frequently used to characterize bound fluxes associated with lattice defects \cite{Willans2010, Vojta2016, Kao2021vacancy, Takahashi2025}.
Assuming $W_p=+1$ for all other bulk plaquettes, we refer to the case $W_{I_j}=-1$ (+1) for all $ j\in\Lambda$ as the bound- (zero-)flux sector.
In each flux sector, the internal plaquettes $W_{+\mu}$ admit four flux configurations constrained by Eq. \eqref{eq: Triple_plaquette}, as shown in Fig. \ref{fig: internal_flux}b. We label these configurations {\textsf{I}}-{\textsf{IV}}, with the subscripts {\textsf{Z}} and {\textsf{B}} denoting the zero-flux and bound-flux sectors, respectively. Corresponding configurations in the zero- and bound-flux sectors are assigned the same label according to the similarity of their local flux patterns.
Among these four configurations, two, denoted as {\textsf{I}}$_{\textsf{Z/B}}$ and {\textsf{IV}}$_{\textsf{Z/B}}$, preserve the mirror-reflection symmetry along the $z$-bond due to their internal flux patterns, whereas the remaining two ({\textsf{II}}$_{\textsf{Z/B}}$ and {\textsf{III}}$_{\textsf{Z/B}}$) break this symmetry.
This property plays a crucial role in the discussion of quadrupole moments of impurity sites. We also note that, even within the zero-flux sector, three of the four internal flux configurations host fluxes, allowing for the possible emergence of MZMs under an applied magnetic field.

\subsection*{Majorana representation for spin operators}
To describe the spin-$1/2$ operators, we use the Kitaev Majorana representation,
\begin{align}
    \hat{S}^\mu = \frac{i}{2} b^\mu c,
\end{align}
where $b^\mu$ and $c$ are Majorana operators satisfying $\{c_j^{},\,c_k^{}\}=2\delta_{jk}$, $\{b_j^\mu,\, b_k^\nu\}=2\delta_{jk}\delta^{\mu\nu}$, and $\{b_j^\mu,\,c_k^{}\}=0$ with $(b_j^\mu)^\dagger = b_j^\mu$ and $c_j^\dagger=c_j^{}$.
Rewriting the bulk Hamiltonian in this representation yields
\begin{align}
    H_0 = \frac{J}{4}
    \sum_{\begin{subarray}{c} j,k\notin \Lambda\\ \langle jk\rangle_\mu \end{subarray}}
    \hat{u}_{j,k}^\mu\,ic_j^{}c_k^{}
\end{align}
with $\hat{u}_{j,k}^\mu\equiv ib_j^\mu b_k^\mu$, and $j$ ($k$) belongs to the white (black) sublattice.
This term describes Majorana hoppings between NN sites.
Similarly, the effective magnetic-field-induced term in the bulk, Eq. \eqref{eq: kappa-term}, takes the form
\begin{align}
    \hat{H}_{\kappa}=\kappa\sum_{(jkl)}u_{j,k}^\mu u_{l,k}^\nu\,ic_j^{}c_l^{},
\end{align}
where $\kappa={3!h^3}/{8(0.065J)^2}$ and the prefactor $3!$ accounts for the number of permutations of the third-order perturbative processes, and a specific ordering of the site triplets $(jkl)$ is assumed in the above expression.
This term describes next-nearest-neighbor (NNN) Majorana hoppings, which open a bulk gap and drive the system into a topologically nontrivial phase.

For spin-3/2 operators we apply the SO(6) Majorana representation \cite{Jin2022}:
\begin{align}
    \begin{split}
       \hat{\Theta}_j^x = \frac{i}{2}\beta_j^x[\gamma_j^{xyz}-(\gamma_j^z - \sqrt{3}\gamma_j^x)],\quad
        \hat{\Theta}_j^y = \frac{i}{2}\beta_j^y[\gamma_j^{xyz}-(\gamma_j^z + \sqrt{3}\gamma_j^x)],\quad
        \hat{\Theta}_j^z = \frac{i}{2}\beta_j^z[\gamma_j^{xyz}+2\gamma_j^z],  
    \end{split}
\label{eq: SO(6)_Majorana}
\end{align}
where $\gamma_j^{xyz}\equiv -i\gamma_j^x\gamma_j^y\gamma_j^z$.
The six Majorana operators $\beta_j^\mu$ and $\gamma_j^\mu$ with $\mu=x,y,z$ act on the impurity site $j\in\Lambda$ [as shown in Fig. \ref{fig: internal_flux}c] and obey the anti-commutation relations as well as the self-conjugate condition.
The operators $\beta_j^\mu$ serve as the $S=3/2$ analog of the fractionalized $b_j^\mu$ Majorana fermions defined on the $\mu$ bonds of the spin-$1/2$ Kitaev model.
In contrast, the matter Majorana fermions on the $S=3/2$ impurity sites come in three flavors,
$\gamma_j^x,\gamma_j^y,$ and $\gamma_j^z$, rather than the  single-flavor $c$-Majorana fermion on the $S = 1/2$ sites.
The local spin-length constraint, $\sum_\mu|\hat{\Theta}_j^\mu|^2 = 15/4$, is automatically satisfied on each impurity site in this representation.
As in the Majorana representation of spin-$1/2$ operators, the local Hilbert space is enlarged by a factor of two. The physical spin-$3/2$ subspace is therefore obtained by imposing a Majorana fermion parity constraint, as detailed below.

The SIA term on the impurity sites then becomes quadratic in terms of matter Majorana fermions,
\begin{align}
\hat{H}_{\rm{SIA}} = D_z \sum_{j\in\Lambda} \left[-\gamma_j^x\gamma_j^y + \frac54\right],
\label{eq: SIA_Majorana}
\end{align}
which represents Majorana hoppings between different flavors of matter Majoranas on the same impurity site.
The other terms involving spin-3/2 operators, $\hat{H}_\Lambda$ and $\hat{H}_\kappa$ around an impurity, on the other hand, are rewritten as interacting Majorana terms and therefore require a mean-field decomposition.
Details of the mean-field treatment are described in the next section.

The SO(6) representation employed here for the spin-3/2 operators offers a key advantage: each impurity site hosts a single “gauge” Majorana fermion $\beta_j^\mu$ associated with a given bond, rather than multiple flavors \cite{Ma2023}.
As a consequence, a static $Z_2$ gauge field emerges not only in the bulk, described by $u_{j,k}^\mu=\pm1$, but also on the bonds connecting an impurity to its NN sites.
Indeed, these gauge fields commute with the total Hamiltonian,
\begin{align}
    [\hat{u}_{j,k}^\mu,\, \hat{H}_{\rm{total}}]=[\hat{\mathcal{U}}_j^\mu,\, \hat{H}_{\rm{total}}]=0
\end{align}
where $\hat{\mathcal{U}}_j^\mu\equiv i\beta_j^\mu b_{k}^\mu$ with $j\in\Lambda$, $k\notin\Lambda$, and $\langle jk\rangle_\mu$.

The flux operators can be rewritten as products of $Z_2$ gauge fields surrounding a given plaquette.
In fact, the flux value for bulk plaquettes can be written as ${W}_p = {u}_{j_2, j_1}^z {u}_{j_2, j_3}^x {u}_{j_4, j_3}^y {u}_{j_4, j_5}^z {u}_{j_6, j_5}^x {u}_{j_6, j_1}^y$.
Similarly, the impurity-related internal plaquette values are given by
\begin{align}
    \begin{split}
    {W}_{+x} = {u}_{8,7}^z {u}_{8,9}^x {\mathcal{U}}_{0}^y {\mathcal{U}}_{0}^z {u}_{6, 5}^x {u}_{6, 7}^y,\quad
    {W}_{+y} = {\mathcal{U}}_{0}^z{\mathcal{U}}_{0}^x{u}_{2,1}^y{u}_{2,3}^z{u}_{4,3}^x{u}_{4,5}^y,\quad
    {W}_{+z} = {u}_{10,9}^z {u}_{10,11}^x {u}_{12,11}^y {u}_{12,1}^z {\mathcal{U}}_{0}^x {\mathcal{U}}_{0}^y.
    \end{split}
\end{align}
The impurity-centered triple-plaquette value is then evaluated as ${W}_{I_0}=\prod_{\langle jk\rangle_\mu\in \partial I_0}{u}_{j,k}^\mu$ with $0\in \Lambda$.
Here we use the identities $\sigma^\mu=-i\epsilon^{\mu\nu\rho}\beta^\nu\beta^\rho/2$ for the pseudo-spin operators of a spin-3/2 impurity \cite{Natori2016}.

Both representations enlarge the local Hilbert space, and therefore one must impose the constraint $\hat{\mathcal{D}}\overset{!}=1$ in order to evaluate physical observables after projection onto the physical subspace.
For a spin-1/2 site, the local constraint is given by $\hat{D}\equiv b^xb^yb^zc\overset{!}=1$, while for a spin-3/2 site it reads $\hat{\Delta} \equiv i\beta^x\beta^y\beta^z\gamma^x\gamma^y\gamma^z\overset{!}=1$.
Upon imposing these constraints, one recovers the correct commutation relations for spin operators.

The projection operator that maps states from the extended Majorana Hilbert space back to the physical subspace can be defined as follows \cite{Pedrocchi2011}:
\begin{align}
\label{eq: proj_op}
\hat{\mathcal{P}}=\prod_j\frac{1 + \hat{\mathcal{D}}_j}{2} = \hat{\mathcal{G}}\cdot\hat{\mathcal{P}}_0,~~{\rm{with}}~~\hat{\mathcal{P}}_0\equiv \frac{1 + \prod_j\hat{\mathcal{D}}_j}{2},
\end{align}
where $\hat{\mathcal{D}}_j$ is given by $\hat{D}_j$ for a spin-1/2 site and by $\hat{\Delta}_j$ for a spin-3/2 site, depending on $j$.
The operator $\hat{\mathcal{G}}$ collects all gauge equivalent sectors satisfying ${\hat{\mathcal{G}}}^2=\hat{\mathcal{G}}$.
Under periodic boundary conditions, the product can be rewritten as
\begin{align}
    \prod_j\hat{\mathcal{D}}_j = \eta\times
    \prod_{\begin{subarray}{c}
        j,k\notin \Lambda,\\
        \langle jk\rangle_\mu
    \end{subarray}}
    u_{j,k}^\mu 
    \times \prod_{j\in\Lambda, \,\mu}\mathcal{U}_j^\mu
    \times \hat{{\pi}}_c,
\end{align}
where $\eta=\pm1$ is a sign factor determined by the reordering of Majorana operators.
The operator $\hat{\pi}_c=\pm1$ denotes the matter-fermion parity sector, which is obtained from the Schur decomposition of the Majorana hopping matrix \cite{Nasu2024_review}.

Regarding the projection onto the physical subspace, we numerically confirmed that, for the reflection-symmetric clusters (defined below) and the parameter sets studied in this work, the physical ground state belongs to the even matter-fermion parity sector, $\hat{\pi}_c=+1$, consistent with the reordering factor $\eta$ and the number of gauge-flipped bonds. The physical ground-state energy is therefore obtained by summing all negative-energy eigenvalues. We further verified, using a projection formalism of Ref.~\cite{Udagawa2018}, that the projected observables coincide with the unprojected ones. We therefore omit the explicit projection in the calculations below.

For completeness, we note that reflection-asymmetric clusters, such as that shown in Fig.~\ref{fig: L32_without_reflection}(a), can instead realize the odd-parity sector, $\hat{\pi}_c=-1$. In this case, both the physical ground-state energy and the quadrupole moments are evaluated with the explicit projection onto the physical subspace.

\begin{figure}[tb]
    \centering
    \includegraphics[width=0.5\linewidth]{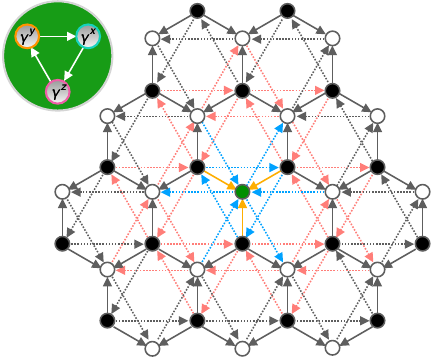}
    \caption{{\textbf{Majorana hopping lattice around an impurity.}}
     Each arrow indicates the positive hopping direction.
    Solid (dashed) lines correspond to NN (NNN) hoppings, respectively.
    Orange and blue arrows require a mean-field decomposition.
    Red arrows do not require such a decomposition; instead, the associated perturbative processes involve internal flux configurations, leading to an enhanced amplitude compared with that in the bulk.
    }
    \label{fig: MF_Majorana}
\end{figure}

\subsection*{Mean-field decomposition}
\label{sec: MF}
We employ a mean-field decomposition technique to analyze the interacting Majorana terms originating from terms $\hat{H}_\Lambda$ and $\hat{H}_\kappa$.
Since both contributions contain terms localized at the impurity sites, we introduce a set of real-space mean-field parameters defined separately for each impurity site.

Our mean-field decomposition involves only operators composed of matter fermions, with the static gauge fields around each impurity site treated exactly and left unrenormalized.
For the mean-field parameters defined on the impurity site $j\in\Lambda$, we have:
\begin{align}
    \tau_j^x \equiv \langle i\gamma_j^y\gamma_j^z\rangle,\quad
    \tau_j^y \equiv \langle i\gamma_j^z\gamma_j^x\rangle,\quad
    \tau_j^z \equiv \langle i\gamma_j^x\gamma_j^y\rangle,
    \label{eq: MF_tau}
\end{align}
which describes pair-condensations between three flavors of matter Majoranas within the same impurity site.
Additionally, we have other mean-fields that represent inter-site couplings:
\begin{align}
    \mathcal{T}_{j,k}^{\mu}\equiv \langle i\gamma_j^\mu c_{k}^{} \rangle\quad (j\in\Lambda,~k\in \partial I_j),\qquad 
    t_{k,l}\equiv \langle i c_{k}^{}c_{l}^{} \rangle\quad (k,l\in \partial I_j).
    \label{eq: MF_T}
\end{align}
Here, $\tau_j^\mu$ and $\mathcal{T}_{j,k}^\mu$ are introduced through the mean-field decomposition of terms such as $i\gamma_j^{xyz}c_k^{}$, which appear from the Kitaev coupling between the impurity and its NN spin-$1/2$ sites, $\hat{\Theta}_j^\mu \hat{S}_{k}^\mu$.
In contrast, $t_{k,l}$ originates from the $\kappa$ term around the impurity.
It is important to note that we do not include any mean-field coupling between the matter and gauge Majoranas.
This is consistent with restricting the Hamiltonian to terms that preserve the local flux operators $\hat{W}_p$, so that the $Z_2$ gauge fields remain static.
Details of mean-field decomposition are presented in the Methods section.

\subsection*{Quadrupole moments}
\label{sec: quadrupole_moment}
Having established the mean-field framework, we now analyze local observables associated with the impurity sites.
 The pseudo-orbital operators $\hat T^\mu$ introduced in Eq. \eqref{Tdefin} represent the multipolar degrees of freedom of the spin-$3/2$ impurity. 
Their expectation values are evaluated in the self-consistent mean-field ground state.
In particular, we focus on the symmetric rank-2 quadrupole moments, which are the lowest-order time-reversal-even observables capable of distinguishing different flux sectors and revealing the local symmetry of the impurity state. 
The corresponding quadrupole components are defined as
\begin{align}
    {Q}^{x}\equiv\langle \hat{T}^x\rangle = -\tau^x, \quad {Q}^{z}\equiv\langle \hat{T}^z\rangle = -\tau^z.
\end{align}
These symmetric quadrupole components ${Q}^{x}$ and ${Q}^{z}$ can acquire finite expectation values, regardless of the flux sector.

In addition to  $Q^x$ and $Q^z$, one may also define the off-diagonal quadrupole components $Q^{\mu\nu}\propto \hat{\Theta}^\mu\hat{\Theta}^\nu + \hat{\Theta}^\nu\hat{\Theta}^\mu$ with $(\mu,\nu)=(x,y), (y,z),$ and $(z, x)$.
However, these components vanish in the self-consistent mean-field ground state. To see this, we express the corresponding operators in the SO(6) Majorana representation and project onto the physical Hilbert space using the local constraint $\hat{\Delta}=1$. Under this projection, they reduce to bilinears involving both gauge and matter Majorana fermions,
\begin{align}
    \hat{Q}^{\mu\nu}\propto \beta^\mu\beta^\nu \gamma^z\gamma^x\propto\hat{\Delta}\,i\beta^\rho\gamma^y
\end{align}
with $(\mu,\nu,\rho)=(x,y,z),\, (y,z,x),$ and $(z,x,y)$.
Since our mean-field ansatz does not include hybridization between gauge Majorana fermions $\beta^\rho$ and matter Majorana fermions $\gamma^\mu$, expectation values of the form $\langle i\,\beta^\rho \gamma^y\rangle$ vanish. 
Consequently, $Q^{\mu\nu}=0$ for these components.

Symmetries constrain the possible values of $(Q^x, Q^z)$ and the associated mean-field parameters $\tau^\mu$.
Under TRS,  the impurity spin operators must transform as $\mathscr{T}\hat{\Theta}^\mu\mathscr{T}^{-1}=-\hat{\Theta}^\mu$. 
In the SO(6) Majorana representation, this condition determines how the Majorana fermions transform under $\mathscr{T}$, leading to \cite{Natori2023}
\begin{align}
    \mathscr{T}\bm{\beta}_j\mathscr{T}^{-1} = (\beta_j^x,\, \beta_j^y,\, \beta_j^z),~~
    \mathscr{T}\bm{\gamma}_j\mathscr{T}^{-1} = (\gamma_j^x,\, -\gamma_j^y,\, \gamma_j^z).
\end{align}
With these, one finds that $Q^\mu=-\tau^\mu$ for $\mu=x,z$ are even under time reversal and may acquire finite expectation values in a $\mathscr{T}$-symmetric state. 
By contrast, $\tau^y$ changes sign under $\mathscr{T}$ and therefore corresponds to a symmetric octupolar degree of freedom, which is forbidden in a $\mathscr{T}$-symmetric system \cite{Carlene2020}.

In addition, when a given cluster preserves global reflection symmetry parallel to the 
$z$-bonds, one necessarily finds $Q^x =0$ \cite{Natori2023}.
This statement holds even for nonuniform systems in the presence of impurities and/or nontrivial flux patterns.
For the minimal cluster with a single, centrally located impurity (Fig. \ref{fig: internal_flux}b), two allowed internal flux patterns preserve reflection symmetry 
 in both the zero- and bound-flux sectors, resulting in states with $Q^x=0$.
This implies that the local quadrupole moment $Q_j^x$ with $j\in \Lambda$ 
 can serve as a marker of the internal flux configuration at an impurity site.
 While reflection symmetry is generally broken for random, multi-impurity configurations, we can strategically arrange impurities to preserve this symmetry (Fig. \ref{fig: L32}a). This allows us to theoretically explore the connection between quadrupole moments and the ground-state flux sector.

\subsection*{Local probe for the quadrupole moment}
Here we propose a potential probe for the quadrupole moment originating from the spin-3/2 impurity. The physical setup uses tunneling spectroscopy to probe the spin-spin correlation function, in which a monolayer Kitaev spin-liquid material is placed between a metallic STM tip and a metallic substrate \cite{Knolle2020, Konig2020, Udagawa2021, Bauer2023, Takahashi2023, KaoPRL2024, Zhang2025_PRB, Zhang2025_npj}. The spin-dependent part of the  differential conductance can be written as
\begin{align}
\frac{\mathrm{d}I}{\mathrm{d}V} \sim \sum_{ik}T(\mathbf{r}-\mathbf{r}_i)T(\mathbf{r}-\mathbf{r}_k) \sum_{\mu\nu}w_{\mu\nu}\int_0^{eV}\mathrm{d}\omega\,S^{\mu\nu}_{ik}(\omega),
\end{align}
where $S^{\mu\nu}_{ik}(\omega)$ is the dynamical spin-spin correlation function of the magnetic system. We assume that the STM tip is positioned directly above the spin-3/2 impurity at site $j$ and is atomically sharp, such that the tunneling amplitude becomes $T(\mathbf{r}-\mathbf{r}_i)T(\mathbf{r}-\mathbf{r}_k) \sim \delta_{ij}\delta_{kj}$ and the response is dominated by the on-site spin-spin correlation. In addition, we consider a spin-polarized setup such that the electrons near the Fermi level in the tip and substrate are fully aligned in spin \cite{Knolle2020}. Therefore, the spin-weight function $w_{\mu\nu}$ can be simplified as
\begin{align}
w_{\mu\nu} = \sum_{\sigma\sigma'} D_{\mathrm{tip}, \sigma}D_{\mathrm{sub},\sigma'}{\hat{\tau}}^{\mu}_{\sigma'\sigma}{\hat{\tau}}^{\nu}_{\sigma\sigma'} \sim \delta_{\mu,z}\delta_{\nu, z}D_{\mathrm{tip},\uparrow}D_{\mathrm{sub},\uparrow},
\end{align}
where $D_{\mathrm{tip}, \sigma}$ and $D_{\mathrm{sub},\sigma'}$ are the spin-dependent electron density of states near the Fermi level for the tip and the substrate, and ${\hat{\tau}}^{\mu}$ and ${\hat{\tau}}^{\nu}$ are Pauli matrices. 

The above setup shows that the spin-dependent differential conductance is proportional to the frequency integral over the $zz$ component of the on-site spin-spin correlation of the impurity. Furthermore, we consider a bias voltage that is above the highest magnetic excitation energy scale $\omega_{\mathrm{max}}$ but still much smaller than the charge gap of the Mott insulator, $\omega_{\mathrm{max}} \lesssim eV \ll \Delta_{\mathrm{charge}}$. 
By using the spectral sum rule, the response is proportional to the equal-time spin-spin correlation function, 
and therefore directly measures the quadrupole moment of the impurity spin:
\begin{align}
\left.\frac{\mathrm{d}I}{\mathrm{d}V}\right\vert_{\mathrm{polarized}} \sim \int_{-\infty}^{\infty} \mathrm{d}\omega\, S^{zz}_{jj}(\omega) \sim \langle \hat{\Theta}^z_j(0)\hat{\Theta}^z_j(0) \rangle = Q^z_j +\frac{5}{4}. 
\end{align}
Note that the above result is based on the zero-temperature formalism where $S^{zz}_{jj}(\omega <0) = 0$, such that only the positive bias voltage is needed. Also, the spin-polarized STM setup is crucial for detecting the quadrupole moment because the unpolarized STM setup simply gives a constant $\langle (\hat{\Theta}^x_j)^2\rangle + \langle (\hat{\Theta}^y_j)^2\rangle + \langle (\hat{\Theta}^z_j)^2\rangle = 15/4$.

\subsection*{Zero-field results}
\begin{figure}[b]
    \centering
    \includegraphics[width=\linewidth]{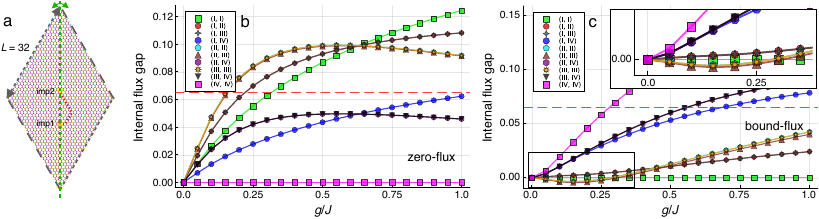}
    \caption{{\textbf{Internal flux excitations in a $L=32$ cluster.}}
    {\textbf{a}} A finite-size cluster with two impurities that preserve reflection symmetry.
    Two impurities are located on different sublattices.
    Yellow plaquettes indicate the presence of $\pi$-fluxes, which corresponds to the internal flux configuration with pattern ({\textsf{IV}}$_{\textsf{B}}$, {\textsf{IV}}$_{\textsf{B}}$) in the bound-flux sector.
    Red bonds represent gauge-flipped bonds that realize the corresponding flux configurations.
    The green dashed line penetrating the cluster denotes the reflection symmetry arising from the impurity configuration. Periodic boundary conditions are imposed.
    {\textbf{b}}, {\textbf{c}} Flux gaps of the internal flux configurations obtained in the zero-flux and bound-flux sectors of the $L=32$ cluster in {\textbf{a}}, respectively.
    Only nonequivalent combinations of the four internal $Z_2$ flux configurations ({\textsf{I}}$_{\textsf{Z/B}}$--{\textsf{IV}}$_{\textsf{Z/B}}$) on the two impurities are shown, since exchanging ({\textsf{imp1}}, {\textsf{imp2}}) gives the same value due to the reflection symmetry of the impurity configuration as well as the boundary condition.
    Some points still overlap as a consequence of the degeneracy between patterns {\textsf{II}} and {\textsf{III}}. The red dashed line in each panel indicates the bulk flux gap. We set $D_z=0.1$ and $h=0$ in both cases.
    The subscripts of {\textsf{I}}--{\textsf{IV}}, {\textsf{Z}} and {\textsf{B}}, are not explicitly shown in labels to avoid redundancy.
    The inset in {\textbf{c}} corresponds to the small coupling regime.}
    \label{fig: L32}
\end{figure}

Starting from this section, we present the results for a cluster containing two impurities.
Most of the analysis is carried out under the assumption of preserved global reflection symmetry.
First, we discuss internal flux excitations in the zero- and bound-flux sectors, and present the corresponding quadrupole moments of impurity sites.
Unless otherwise stated, we focus on the two-impurity cluster shown in Fig.~\ref{fig: L32}a as a representative example.

As discussed above, each impurity admits four internal $Z_2$ flux configurations ({\textsf{I}}$_{\textsf{Z/B}}$-{\textsf{IV}}$_{\textsf{Z/B}}$), leading to
$4\times4$ possible patterns of internal $Z_2$ flux configurations in the two-impurity cluster in both zero- and bound-flux sectors.
A given internal flux pattern can therefore be specified by a pair of labels, such as ({\textsf{I}}$_{\textsf{Z/B}}$, {\textsf{I}}$_{\textsf{Z/B}}$) through ({\textsf{IV}}$_{\textsf{Z/B}}$, {\textsf{IV}}$_{\textsf{Z/B}}$), 
indicating the configurations at the two impurity sites (\textsf{imp1}, \textsf{imp2}).
For a given flux sector, one can optimize the mean-field parameters for each allowed internal flux pattern and compute the corresponding ground-state energies.
Here, we focus on the lowest-energy internal flux configuration within each flux sector.
Table~\ref{tab:flux_quadrupole_summary} summarizes the notation and physical properties discussed in this section.

\begin{table}[t]
\centering
\caption{Summary of flux configurations, reflection symmetry, and quadrupole moments for $D_z>0$ and at zero field for the $L=32$ cluster with the global reflection symmetry shown in Fig~\ref{fig: L32}a. The internal flux transition point in the bound-flux sector, $g^*$, is proportional to $D_z$.}
\label{tab:flux_quadrupole_summary}
\renewcommand{\arraystretch}{1.2}
\begin{tabular}{|c|c|c|c|}
\hline
\multirow{2}{*}{\makecell{Ground-state \\ flux configuration}}
& \multicolumn{3}{c|}{$W_p=+1$ for bulk plaquettes} \\
\cline{2-4}
& Zero-flux sector
& \multicolumn{2}{c|}{Bound-flux sector} \\
\hline

\makecell{$W_I=W_{+x}W_{+y}W_{+z}$ \\ at both impurities} & $+1$ & \multicolumn{2}{c|}{-1} \\
\hline

\multirow{2}{*}{\makecell{Lowest internal flux patterns \\
at $(\textsf{imp1},\textsf{imp2})$}}
& $(\textsf{IV}_{\textsf{Z}},\textsf{IV}_{\textsf{Z}})$
& \makecell{$(\textsf{II}_{\textsf{B}},\textsf{III}_{\textsf{B}})$,
$(\textsf{III}_{\textsf{B}},\textsf{II}_{\textsf{B}})$}
& $(\textsf{I}_{\textsf{B}},\textsf{I}_{\textsf{B}})$ \\

\makecell{}
& unique
& two-fold degenerate for $g<g^*$
& unique for $g>g^*$ \\
\hline

\makecell{$(W_{+x},W_{+y},W_{+z})$ at $\textsf{imp1}$}
& $(-1,-1,+1)$
& \makecell{$(-1,+1,+1)$, $(+1,-1,+1)$}
& $(-1,-1,-1)$
\\

\makecell{$(W_{+x},W_{+y},W_{+z})$ at $\textsf{imp2}$}
& $(-1,-1,+1)$
& \makecell{$(+1,-1,+1)$, $(-1,+1,+1)$}
& $(-1,-1,-1)$
\\
\hline

Local reflection symmetry
& $\checkmark$
& $\times$
& $\checkmark$
\\
\hline

$(Q_{\textsf{imp1}}^x,Q_{\textsf{imp2}}^x)$
& $(0,0)$
& $(+q,-q)$, $(-q,+q)$
& $(0,0)$
\\

$|Q_{\textsf{imp1}}^z|=|Q_{\textsf{imp2}}^z|$
& relatively large
& relatively small
& relatively small
\\
\hline
\end{tabular}
\end{table}

In the zero-flux sector, we find that the ({\textsf{IV}$_\textsf{Z}$}, {\textsf{IV}}$_\textsf{Z}$) pattern always realizes the lowest energy, protected by a finite {\textit{internal}} flux gap $\Delta_{\rm{int}}$.
The ground-state energy difference between the lowest-energy pattern ({\textsf{IV}}$_\textsf{Z}$, {\textsf{IV}}$_\textsf{Z}$) and other configurations is shown in Fig.~\ref{fig: L32}b.
The figure indicates that the lowest internal flux configuration is unique for finite $g$ and that $\Delta_{\rm{int}}$ depends on the coupling strength $g$.
We also confirm that, in the zero-flux sector, $\Delta_{\rm{int}}$ is always positive for $D_z>0$.

In the bound-flux sector, on the other hand, we observe an internal flux transition as a function of $g/J$.
Taking the ({\textsf{I}}$_\textsf{B}$, {\textsf{I}}$_\textsf{B}$) pattern as the energy reference, we plot the internal flux gaps to several other flux configurations in Fig.~\ref{fig: L32}c.
The figure shows that, for small but finite $g/J$, some values of $\Delta_{\rm{int}}$ become negative, indicating that the lowest internal flux configuration is no longer the ({\textsf{I}}$_\textsf{B}$, {\textsf{I}}$_\textsf{B}$) but instead belongs to the twofold-degenerate sectors, ({\textsf{II}}$_\textsf{B}$, {\textsf{III}}$_\textsf{B}$) and ({\textsf{III}}$_\textsf{B}$, {\textsf{II}}$_\textsf{B}$).
Two other configurations, ({\textsf{II}}$_\textsf{B}$, {\textsf{II}}$_\textsf{B}$) and ({\textsf{III}}$_\textsf{B}$, {\textsf{III}}$_\textsf{B}$), lie slightly higher in energy than the lowest configuration.
The essential difference between the lowest and those slightly higher configurations is as follows:
in the former, each impurity site hosts a pair of positive and negative $Q^x$ values with the same magnitude as $(Q_{\textsf{imp1}}^x, Q_{\textsf{imp2}}^x)=(+q,-q), (-q,+q)$ with $q>0$, resulting in,  $\sum_{j\in\Lambda} Q^x_j=0$.
The latter configurations, on the other hand, exhibit a finite net $Q^x$, like $(Q_{\textsf{imp1}}^x, Q_{\textsf{imp2}}^x)=(+q,+q), (-q,-q)$.
For larger values of $g/J$,
the unique ({\textsf{I}}$_\textsf{B}$, {\textsf{I}}$_\textsf{B}$) pattern emerges as the ground-state flux configuration, 
protected by a finite flux gap.
We find that the internal flux transition point $g^*$ ($\sim0.3J$ in Fig.~\ref{fig: L32}c) is proportional to $D_z$.

\begin{figure*}[t]
    \centering
    \includegraphics[width=\linewidth]{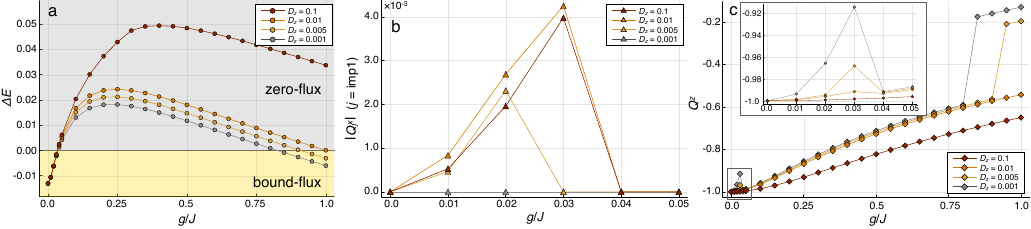}
    \caption{{{\textbf{Ground-state flux sector and associated impurity quadrupole moments.}}}
    All calculations are performed at zero field in the $L=32$ cluster shown in Fig. \ref{fig: L32}a.
    {\textbf{a}} Difference in the ground-state energies between the zero-flux and bound-flux sectors $\Delta E$. 
        {\textbf{b}} Local quadrupole moments of $|Q_j^x|$ with $j=${\textsf{imp1}} for small coupling regime.
        The finite values of $|Q^x_j|$ are numerically robust: for reflection-symmetric impurity configurations and internal flux patterns, where symmetry requires $Q^x=0$, we obtain $|Q^x|\lesssim 10^{-9}$.
    {\textbf{c}} Local quadrupole moments of $Q^z\equiv Q_{\textsf{imp1}}^z=Q_{\textsf{imp2}}^z$. The inset corresponds to the small coupling regime.}
    \label{fig: L32_lowest}
\end{figure*}

Next, we examine which flux sector is energetically favored by comparing the zero- and bound-flux sectors.
Defining the energy difference between the two flux sectors as
\begin{align}
\Delta E \equiv E_{\rm bound} - E_{\rm zero},
\end{align}
we plot $\Delta E$ as a function of $g$ for different values of $D_z$, as shown in Fig.~\ref{fig: L32_lowest}a.
To analyze this behavior, we identify couplings $g_1$ and $g_2$ as the points corresponding to transitions
between the bound- and zero-flux sectors (bound $\rightarrow$ zero and zero $\rightarrow$ bound, respectively), along with the internal flux transition point $g^*$ within the bound-flux sector.

At $g/J=0$, the system is always in the bound-flux sector because the impurities are decoupled from the bulk and effectively act as vacancies \cite{Willans2010}.
For finite but very small $g$, the system still stays in the bound-flux sector; however, the \textit{internal} flux configuration breaks the local reflection symmetry, as discussed in the previous section.
As $g$ is increased, the system undergoes the first flux sector transition from the bound-flux sector to the zero-flux sector. Notably, this transition occurs at $g_1$ and is essentially independent of $D_z$.
For small $D_z$, a second transition appears at a larger coupling $g_2$, where the system returns to the bound-flux sector while maintaining the local reflection symmetry, leading to a reentrant behavior.

This reentrance is expected to be limited to the small-$D_z$ regime for the following reason.
For $D_z>0$, the spin-1/2 doublet at an impurity site lies lower in energy than the spin-3/2 states. Consequently, in the limit $D_z \rightarrow \infty$ the impurity effectively reduces to a spin-1/2 degree of freedom, with only a renormalization of the coupling strengths \cite{Natori2023}.
This limit corresponds to the quasivacancy case \cite{Kao2021vacancy}, which does not show reentrant behavior.

The ordering of the characteristic points depends on the magnitude of $D_z$: for small but finite $D_z$, $g^*<g_1<g_2$ whereas for sufficiently large $D_z$, $g_1<g^*\ll g_2$.
In the latter case, the internal transition at $g^*$ is not directly observable because the bound-flux sector is energetically higher than the zero-flux sector around this point.

With the lowest-energy flux configuration as a function of $g/J$, we now discuss the local quadrupole moments ($Q_j^x, Q_j^z$) with $j=$ \textsf{imp1}, \textsf{imp2}.
Importantly, we find that $Q^z\equiv Q_{\textsf{imp1}}^z=Q_{\textsf{imp2}}^z$ (due to the reflection symmetry of the impurity configuration)
serves as a direct probe of flux sector transitions as shown in Fig.~\ref{fig: L32_lowest}c: it exhibits discontinuous jumps at the transition points $g_1$ and $g_2$ (if present), regardless of $D_z$.  In addition, we note that $|Q^z|$ is generally small in the bound-flux sector compared to that in the zero-flux sector.
 This can be understood from two complementary perspectives:
(i) Within the mean-field description, symmetry allows the Majorana operator $\gamma^y$ to couple only to $\gamma^x$ on the impurity site,
 whereas $\gamma^x$ couples to both $\gamma^y$ on the impurity and to the neighboring host $c$-Majoranas (see Fig. \ref{fig: Lieb_theorem}).
Consequently, the local quadrupole moment $Q^z=-i\gamma^x\gamma^y$ is determined by the competition between the intra-site $\gamma^x$-$\gamma^y$ pairing and the impurity-host hybridization of $\gamma^x$. 
A positive SIA $D_z$ favors a large local $\gamma^x$-$\gamma^y$ correlation, stabilizing a large negative $Q^z$. 
When a $Z_2$ flux is bound to the impurity, however, the local $Z_2$-field configuration modifies the coupling of $\gamma^x$ to neighboring $c$-Majorana fermions.
This redistributes the spectral weight of $\gamma^x$ from the impurity to the host, suppressing the local bilinear and thereby reducing $|Q^z|$.
(ii) From a spin perspective, $D_z>0$ favors the $|m_z|=1/2$ states of the spin-$3/2$ impurity, leading to a large negative quadrupole moment. 
In the bound-flux sector, the impurity becomes more strongly entangled with the surrounding bulk spins through the increased bond-dependent interactions
$g$. Based on the mean-field calculations, this enhanced entanglement increases the admixture of $|m_z|=3/2$ components into the ground state.  This admixture effectively renormalizes the anisotropy $D_z$ to a smaller value and shifts $Q^z$ toward zero, thereby reducing $|Q^z|$ compared to the zero-flux sector.

By contrast, $Q_j^x$ is a probe of the \textit{internal} flux configuration and the associated local reflection symmetry.
When $g_1<g^*$ for large $D_z$, the flux sector transition at $g_1$ is accompanied by a jump of $|Q_j^x|$ from a finite value to zero, because the system changes from a symmetry-breaking bound-flux state with patterns ({\textsf{II}}$_{\rm{B}}$, {\textsf{III}}$_\textsf{B}$) and ({\textsf{III}}$_{\rm{B}}$, {\textsf{II}}$_\textsf{B}$) at $g<g_1$ to a symmetry-preserving zero-flux state with pattern ({\textsf{IV}}$_\textsf{Z}$, {\textsf{IV}}$_\textsf{Z}$) at $g>g_1$.
This behavior is observed for $D_z=0.01$ and $D_z = 0.1$ as shown in Fig. \ref{fig: L32_lowest}b.
When $g^*<g_1$, the nonzero-to-zero jump in $|Q_j^x|$ instead occurs at $g^*$ within the bound-flux sector, where the internal-flux transition restores the local reflection symmetry. Consequently, this jump does not coincide with the flux sector transition at $g_1$.
This corresponds to the case $D_z=0.005$ in Fig. \ref{fig: L32_lowest}b.
Finally, at sufficiently small $D_z$, $g^*$ is not resolved within our scanned $g$ grid, and accordingly, we do not observe a clear nonzero-to-zero jump in $|Q_j^x|$ as in the case $D_z=0.001$.
In short, $Q_j^z$ identifies the flux sector transitions, $g_1$ and $g_2$, while $Q_j^x$ identifies the internal reflection symmetry for each flux sector.

\subsection*{Results in a finite magnetic field}
\label{sec: result_nonzerofield}
\begin{figure}[t]
    \centering
    \includegraphics[width=\linewidth]{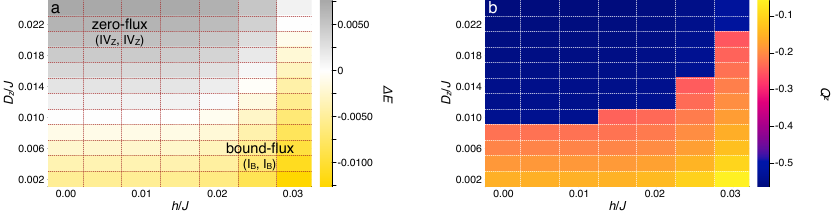}
    \caption{{\textbf{Phase diagrams in the parameter space of $(h/J,D_z/J)$ at $g/J=1$.}}
    Both brown and white dotted lines indicate the boundaries of the pixels.
    {\textbf{a}} The heatmap of the energy difference $\Delta E$.
    We use the same cluster as shown in Fig.~\ref{fig: L32}a.
    {\textbf{b}} The heatmap of the local quadrupole moment $Q_j^z$ in the lowest flux sector} for the same cluster and parameter points.
    \label{fig: phase_diagram}
\end{figure}
 
 We here introduce the effective $\kappa$ term, which gaps out the bulk Majorana spectrum and gives rise to localized MZMs in the bound-flux sector.
Furthermore, we examine the behavior of quadrupole moments and their correlations, both in the presence and absence of these zero modes.

The effective $\kappa$ term in Eq. \eqref{eq: kappa-term} induces Majorana hoppings between NNN sites, with the magnitude depending on the local flux excitation gap.
Since the internal flux gap around an impurity exhibits complicated parameter dependence,  $\Delta_{\rm{int}}(g, \, D_z,\, \{\mathcal{U}^{\mu}\}, \{u^{\mu}\})$, 
as shown in Figs. \ref{fig: L32}b and c, evaluating the hopping amplitudes as explicit functions of these parameters becomes cumbersome.
Instead, for simplicity, we assume that $\Delta_{\rm{int}}$ takes a constant value of $0.02J$, which is at least valid for $g/J=1$ with $0<D_z\leq0.1$ for the cluster of Fig. \ref{fig: L32}a, as well as for the similar clusters that we examined.
This results in an enhancement of the NNN hopping amplitudes compared with those in the bulk.
We apply this amplification to all hoppings around each impurity, indicated by the blue or red dashed arrows in Fig. \ref{fig: MF_Majorana}.
We here note that the parameter dependence of $\Delta_{\rm{int}}$ appears as renormalization of $\kappa$ or the field strength $h$ in phase diagrams.

We evaluate the stability of the lowest flux sector at $g/J=1$ for various values of $D_z$ in the presence of a magnetic field. The resulting phase diagram is shown in Fig.~\ref{fig: phase_diagram}a.
 In the absence of the magnetic field, there is a critical $D_z$ that determines which of the bound-flux and zero-flux sectors is the ground state, as discussed in Fig.~\ref{fig: L32_lowest}a. In the presence of fields, the boundary of the two sectors bends upward in Fig.~\ref{fig: phase_diagram}a, indicating the stabilization of the bound-flux sector and an increase in the critical value of $D_z$. 
 This gives rise to another possible transition from the zero-flux sector to the bound-flux sector driven by external magnetic fields at moderate $D_z$.

\begin{figure*}
    \centering
    \includegraphics[width=\linewidth]{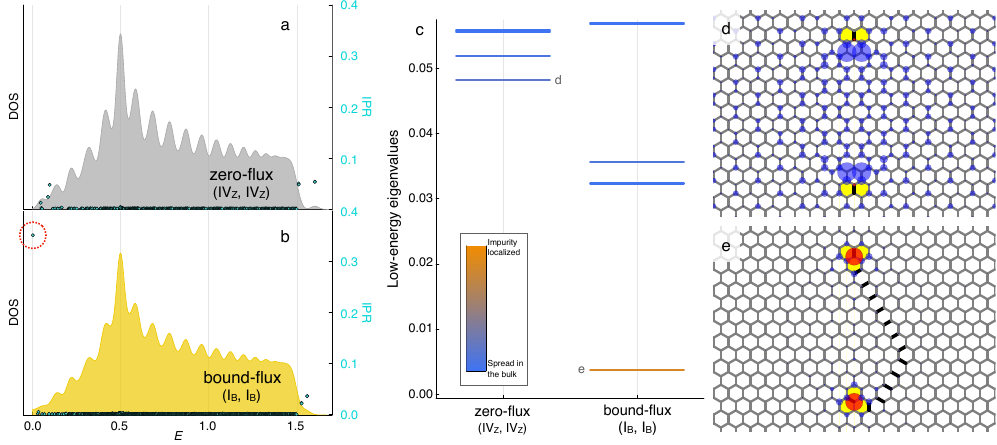}
    \caption{{\textbf{Eigenstates in the zero- and bound-flux sectors.}}
    The cluster shown in Fig. \ref{fig: L32}a is used.
    The parameters are $(g, D_z, h) = (1, 0.006, 0.02)$ in units of $J$. For these parameters, the bound-flux sector has the lowest energy in all panels shown here.
    {\textbf{a}}, {\textbf{b}} DOS and IPR in the zero-flux and bound-flux sectors. Light blue dots indicate the IPR for all states with positive eigenvalues.
    A localized MZM is highlighted with a red dashed circle in  {\textbf{b}}.
    {\textbf{c}} Low-energy positive eigenvalues in the two flux sectors. The color of each line, ranging from blue to orange, indicates the fraction of the corresponding wavefunction localized on the impurity $\gamma$-Majorana sites, $0\leq\sum_{j\in\Lambda}\sum_{\mu=x,y,z} |\psi_{\gamma_j^\mu}(E_i)|^2\leq1$. The lowest-energy modes highlighted here correspond to the real-space wavefunctions shown in {\textbf{d}} and {\textbf{e}} for the zero-flux and bound-flux sectors, respectively.
    {\textbf{d}} Real-space distribution of the wavefunction for the lowest-energy mode in the zero-flux sector. The radius of each circle represents the local weight of the wavefunction at each site. At non-impurity sites, this weight is given by $|\psi_{j}(E_1)|^2$. At impurity site $j\in\Lambda$, the red circle represents the summed weight $\sum_{\mu=x,y,z} |\psi_{\gamma_j^\mu}(E_1)|^2$.
    For visibility, the blue circles at non-impurity sites are enlarged by a factor of 100 relative to the red circles at the impurity sites.
    {\textbf{e}} Same as {\textbf{d}}, but shown for the bound-flux sector, where
     $E_1=E_{\rm{MZM}}$.}
    \label{fig: low_energy_dist}
\end{figure*}

As discussed above, a discontinuous jump in the local quadrupole moment $Q_j^z$ occurs when the lowest flux sector transitions between the bound-flux sector with pattern (\textsf{I}$_{\textsf{B}}$, \textsf{I}$_{\textsf{B}}$) and the zero-flux sector with pattern (\textsf{IV}$_{\textsf{Z}}$, \textsf{IV}$_{\textsf{Z}}$).
Fig.~\ref{fig: phase_diagram}b shows a heatmap of $Q_j^z$ at $j=$ {\textsf{imp1}} for the same cluster, clearly highlighting the distinct flux sectors in the parameter space ($h/J$, $D_z/J$), where $h = |\bm{h}|$ is the magnitude of the field strength along [111].
We note again that $Q_j^x$ vanishes in both flux sectors, since both the global and local reflection symmetries are preserved by the impurity configuration in the cluster and by the lowest-energy flux configurations in zero- and bound-flux sectors at $g/J=1$.

Finally, we examine how the stability of the bound-flux sector depends on the impurity separation. 
For a fixed cluster size $L=32$, we consider several impurity configurations that preserve the global reflection symmetry and evaluate $\Delta E$ at $g/J=1$. 
For larger distance between the two impurities, we find that the zero-flux sector is more favored at finite $D_z$, but a transition into the bound-flux sector can be induced by increasing the magnetic fields.

It is important to analyze the Majorana spectra in both flux sectors separately, since we find that the lowest-energy internal flux configuration in each flux sector hosts a nontrivial $Z_2$ flux pattern.
Figs. \ref{fig: low_energy_dist}a and b show the density of states (DOS) for each flux sector evaluated in its lowest-energy internal flux configuration.
A crucial difference between the two sectors lies in the presence or absence of MZMs.
In the bound-flux sector, we observe a pair of MZMs, or more precisely Majorana quasi-zero modes due to the finite overlap between impurities, protected by the bulk Majorana gap.
This is expected since each impurity site hosts a composite (triple-plaquette) flux, $W_I=-1$, and each such flux binds a MZM.
In contrast, no zero modes appear within the finite bulk gap in the zero-flux sector.
This can be understood from the internal-flux structure of this sector: when two internal fluxes are present per impurity, the corresponding MZMs can hybridize with each other and split away from zero energy.

To characterize the low-energy eigenmodes, we denote by $\psi_j(E_i)$ the wavefunction of the eigenstate with energy $E_i$, where $j$ labels the Majorana site. The spatial extent of each eigenmode is quantified by the inverse participation ratio (IPR), $\mathrm{IPR}(E_i)=\sum_j |\psi_j(E_i)|^4$, where a larger IPR indicates a more localized eigenmode. As shown in 
Fig.~\ref{fig: low_energy_dist}b, the MZM in the bound-flux sector exhibits a significantly larger IPR than the other low-energy modes. The real-space distributions of the corresponding wavefunctions, shown in Fig.~\ref{fig: low_energy_dist}d and e, further demonstrate that the zero mode in the bound-flux sector is strongly localized at the impurity sites, whereas the lowest-energy mode in the zero-flux sector remains much more extended.

To examine how the flux sector influences correlations between impurities, we define the equal-time connected quadrupole correlation as
\begin{align}
    C^\mu(\Delta r)\equiv \langle Q_{\textsf{imp1}}^\mu Q_{\textsf{imp2}}^\mu\rangle - \langle Q_{\textsf{imp1}}^\mu\rangle \langle Q_{\textsf{imp2}}^\mu\rangle
\end{align}
with $\mu=x$ and $z$.
Here, $\Delta r$ denotes the relative distance between the two impurities, measured in units of the lattice constant of the unit cell of the honeycomb lattice.
The second term is required only for the $z$ component in order to subtract trivial local moments, as $Q_j^x=0$ in clusters with global reflection symmetry.
\begin{figure}
    \centering
    \includegraphics[width=0.95\linewidth]{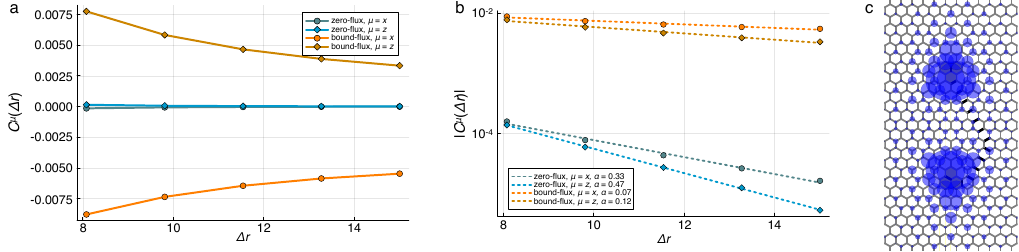}
    \caption{{\textbf{Quadrupole correlations in the zero-flux sector with (\textsf{IV}$_{\textsf{Z}}$, \textsf{IV}$_{\textsf{Z}}$) and the bound-flux sector with (\textsf{I}$_{\textsf{B}}$, \textsf{I}$_{\textsf{B}}$)}}.
    Here $(g,D_z,h)=(1,0.006,0.02)$ in units of $J$ and we use the same finite-size cluster shown in Fig. \ref{fig: L32}a.
    {\textbf{a}} Linear plot of $C^\mu(\Delta r)$ for $\mu=x,z$.
    {\textbf{b}} Linear-log plot of $|C^\mu(\Delta r)|$. Each dashed line shows a fitted exponential function with the decay rate $\alpha$.
    {\textbf{c}} Real-space distribution of the wavefunction for the second lowest-energy mode in the bound-flux sector with (\textsf{I}$_{\textsf{B}}$, \textsf{I}$_{\textsf{B}}$). Details of this plot are the same as Figs. \ref{fig: low_energy_dist}d and e.}
    \label{fig: L32_quadrupole_correlation}
\end{figure}

Figure \ref{fig: L32_quadrupole_correlation} summarizes the distance dependence of the equal-time quadrupole correlations between two impurities.
As shown in Fig. \ref{fig: L32_quadrupole_correlation}a, the correlations decay monotonically with increasing impurity separation $\Delta r$ for both flux sectors and for both $\mu=x$ and $z$ components{, indicating that they are governed by the overlap of the low-energy wavefunctions.} The sign of the correlation depends on the component: while the $z$ component exhibits ferro-quadrupolar correlations, the $x$ component shows antiferro-quadrupolar correlations. This difference originates from the finite positive anisotropy $D_z$, which favors alignment of the $z$-quadrupole moments.

A clear distinction between the two flux sectors appears in the magnitude of the correlations. The correlations in the bound-flux sector are roughly two orders of magnitude larger than those in the zero-flux sector,
indicating that the low-energy impurity-bound states in the bound-flux sector strongly enhance the impurity-induced quadrupole correlations.
The behavior of $C^\mu(\Delta r)$ is notably different from that of the local quadrupole moments in Fig. \ref{fig: phase_diagram}b. This is because the two observables probe different aspects of the impurity response. While $Q_j^\mu$ measures the local quadrupolar polarization, $C^\mu(\Delta r)$ measures nonlocal connected fluctuations between impurities. The much larger correlations in the bound-flux sector therefore indicate not a larger local moment, but a more efficient channel of inter-impurity coupling mediated by impurity-bound states.

To quantify the decay behavior, we fit the absolute value of the correlations to an exponential form $|C^\mu(\Delta r)| \propto e^{-\alpha \Delta r}$, as shown in Fig. \ref{fig: L32_quadrupole_correlation}b.
In the zero-flux sector, where there is a finite energy gap due to the magnetic field, we obtain $0.3 \lesssim \alpha \lesssim 0.5$ for our parameter regime $(0.0<h\leq0.03, 0.002 \leq D_z\leq 0.024)$, corresponding to a correlation length of the order $\xi=\mathcal{O}(1)$.
This short-ranged behavior is consistent with impurity correlations mediated by gapped bulk excitations, for which the decay length is expected to be set by the bulk correlation length and thus to be controlled by the inverse gap scale.

In contrast, the bound-flux sector exhibits much smaller decay  rates, $0.05 \lesssim \alpha \lesssim 0.15$, corresponding to $\xi=\mathcal{O}(10)$. 
This behavior can be understood from the configuration of the local gauge fields, which modify the effective Majorana hopping amplitudes and thereby reconstruct the low-energy spectrum, giving rise to localized excitations such as MZMs (Fig.~\ref{fig: low_energy_dist}c).
The slowly decaying impurity correlations originate from the finite overlap of these localized modes, in contrast to the bulk-mediated correlations in the zero-flux sector.
At the same time, $C^\mu$ is not determined solely by the lowest-energy mode: since it involves two-particle excitations, higher-energy modes also contribute. While the MZM is strongly localized, these higher-energy modes remain more extended as shown in Fig. \ref{fig: L32_quadrupole_correlation}c, which leads to a residual dependence of the correlations on impurity separation even when the projection operator defined in Eq. \eqref{eq: proj_op} is taken into account, unlike in the vacancy case of Ref. \cite{Takahashi2023}.

Once a sufficiently large mass term is introduced by hand to gap out the bulk spectrum, the correlations in the zero-flux sector become numerically negligible, whereas in the bound-flux sector the correlations remain finite, on the order of $\mathcal{O}(10^{-5})$, and exhibit exponential decay.
This contrast also indicates that the correlations in the zero-flux sector are primarily mediated by bulk low-energy modes, while in the bound-flux sector they are sustained by impurity-bound states that survive even when the bulk spectrum is strongly gapped.

\subsection*{Clusters without reflection symmetry}
Our analysis so far has focused on clusters that preserve reflection symmetry due to the impurity configuration. 
Such symmetric arrangements are relevant for candidate materials, such as $\alpha$-RuCl$_3$ \cite{Plumb2014, Baek2017}, with a low concentration of spin-3/2 impurities (e.g., Cr$^{3+}$ \cite{Bastien2019, Lee2023} or Zr$^{3+}$ \cite{Natori2025}), where reflection symmetry is preserved on average even when impurities are randomly distributed.

\begin{figure}[t]
    \centering
    \includegraphics[width=\linewidth]{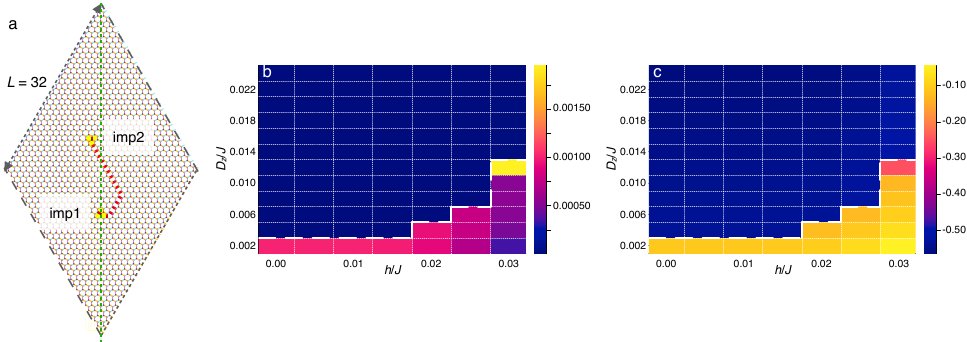}
    \caption{\textbf{Results for a cluster without global reflection symmetry.}
    {\textbf{a}} A cluster without global reflection symmetry.
    {\textbf{b}} The heatmap of $|Q_j^x|$ with $j=${\textsf{imp1}} and $g/J=1$.
     The white thick dashed lines represent the flux sector transition in this cluster confirmed by $\Delta E$.
     Thin white dotted lines indicate the boundaries of the pixels.
    {\textbf{c}} The heatmap of $Q_j^z$ for the same parameter setting.
    }
    \label{fig: L32_without_reflection}
\end{figure}

It is nevertheless instructive to examine the behavior when this symmetry is broken due to a configuration of impurities.  Figure \ref{fig: L32_without_reflection}a shows a cluster geometry in which reflection symmetry is absent. 
We confirm that, even for a cluster without global reflection symmetry, the lowest-energy internal flux configurations at $g/J=1$ are (\textsf{IV}$_{\textsf{Z}}$, \textsf{IV}$_{\textsf{Z}}$) in the zero-flux sector and (\textsf{I}$_{\textsf{B}}$, \textsf{I}$_{\textsf{B}}$) in the bound-flux sector. The corresponding energy difference,
$\Delta E = E_{\rm bound,odd}-E_{\rm zero,even}$,
yields a phase diagram qualitatively similar to that in Fig.~\ref{fig: phase_diagram}(a), where we have used the odd-parity ground-state energy in the bound-flux sector, as discussed at the end of the “Majorana representation for spin operators” subsection.

Based on this result, we further evaluate the local quadrupole moments $(Q_j^x,Q_j^z)$ at the impurity sites ($j=\textsf{imp1},\textsf{imp2}$) by explicitly projecting onto the physical subspace in each flux sector using the projection formalism of Ref.~\cite{Udagawa2018}.
Figs. \ref{fig: L32_without_reflection}b and c show heatmaps of the quadrupole moments $(|Q_j^x|, Q_j^z)$ with $j=$ \textsf{imp1} 
in this cluster.
We find that both quadrupole moments remain sensitive to the lowest flux sector; however, the magnitude of $|Q_j^x|$ is significantly smaller than that of the $z$-component.

\subsection*{Stability of $\pi$-flux and Lieb's conjecture }
\label{sec: Lieb's conjecture}
We can make a plausible connection between the stability of $\pi$-fluxes bound to spin-3/2 impurities and Lieb's conjecture on flux configurations.
The conjecture originally states that, for a half-filled band of electrons hopping on a planar bipartite lattice, the energy-minimizing magnetic flux through each elementary plaquette is $\pi$ if the number of sites in the loop is congruent to $0$ modulo $4$, and $0$ if it is congruent to $2$ modulo $4$, as in square and hexagonal plaquettes, respectively \cite{Lieb1992}.
The magnetic flux $\phi$ is defined as the phase acquired by the wavefunction upon traversing a closed loop, given by
    \begin{align}
        e^{i\phi}=\prod_{{\rm{(B,A)}}\in\rm{closed\,loop}}{\rm{sign}}(t_{\rm{BA}}).
    \end{align}
Here, A and B denote two sublattices, respectively, and $t_{\rm{BA}}$ represents the hopping amplitude between these two sublattices.
 Examples in the QSL literature include the $\pi$-flux phase proposed for cuprate parent compounds \cite{Laughlin1988, Hasegawa1989}, the staggered $[0,\pi]$ flux ansatz in the $J_1$-$J_2$ triangular-lattice model \cite{Iqbal2016, Willsher2025}, and the zero-flux ground state of the Kitaev spin liquid in the absence of lattice defects \cite{Kitaev2006}.

We are therefore interested in the structure of the Majorana hopping Hamiltonian obtained after mean-field optimization in the absence of a magnetic field.
In the bulk, Majorana hoppings occur only between different sublattices of the honeycomb lattice.
Around an impurity, however, the optimized hopping network can become more complex, since three $\gamma$ Majorana sites are present and can hybridize among themselves through $\tau^\mu$ and with their nearest neighbors through $\mathcal{T}_{j,k}^\mu$.

The optimized zero-field state must satisfy constraints on the mean-field parameters,
\begin{align}
\tau_j^y = \langle i\gamma_j^z\gamma_j^x\rangle=0,\quad
\mathcal{T}_{j,j+\mu}^{y} = \langle i\gamma_j^y c_{j+\mu}^{}\rangle=0
\label{eq: tau_y}
\end{align}
with $\mu=x,y,$ and $z$ for the second constraint.
The first constraint follows from  TRS of the system, while the second arises because $\mathcal{T}_{j,j+\mu}^{y}$ corresponds to the adjoint component of $\tau^y$ in the mean-field decomposition, as shown in Eq. \eqref{eq: MF_imp_coup}.
Despite these constraints, the optimized Majorana hopping lattice can be nonplanar, which makes a direct application of Lieb’s conjecture nontrivial in general.

\begin{figure}[t]
    \centering
    \includegraphics[width=0.6\linewidth]{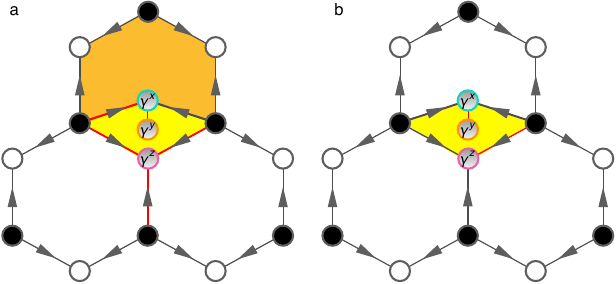}
    \caption{\textbf{Mean-field optimized Majorana hopping lattice around an impurity in a cluster preserving the global reflection symmetry.}
    Each bond represents a finite hopping amplitude, with its thickness indicating the magnitude. Arrows define the positive hopping direction. Black bonds follow the convention of positive hopping, while red bonds indicate negative hopping amplitudes.
    Parameters are set as $(g/J,\, D_z/J,\,h/J)=(1,\,0.006, \,0)$ in the cluster shown in Fig. \ref{fig: L32}a.
    {\textbf{a}} The zero-flux sector with (\textsf{IV}$_{\textsf{Z}}$, \textsf{IV}$_{\textsf{Z}}$). Two emergent $\pi$-fluxes are present, indicated by the orange and yellow plaquettes.
    {\textbf{b}} The bound-flux sector with (\textsf{I}$_{\textsf{B}}$, \textsf{I}$_{\textsf{B}}$). A single $\pi$-flux is localized on the emergent square plaquette.
    }
    \label{fig: Lieb_theorem}
\end{figure}

Here we can simplify the discussion by exploiting the fact that, in clusters with global reflection symmetry at $g/J=1$, the lowest internal flux configurations also respect local reflection symmetry in both flux sectors. This symmetry implies $Q_j^x\sim i\gamma_j^y\gamma_j^z=0$ with $j\in\Lambda$.
Under this condition, we find that the mean-field optimized state in each flux sector further satisfies $\mathcal{T}_{j,j+z}^x=\langle i\gamma_j^x c_{j+z}^{}\rangle =0$.
The vanishing of these Majorana hopping amplitudes deforms the Majorana hopping lattice around an impurity into a planar graph as shown in Fig. \ref{fig: Lieb_theorem}.
In this planar structure, the Majorana mode $\gamma^x$ is connected to $j+x$ and $j+y$ sites, $\gamma^z$ is connected to all NN sites, while $\gamma^y$ is connected only to $\gamma^x$ and does not form any closed circuits.
As a result, all sites can be divided into two distinct sublattices, and all Majorana hopping amplitudes are finite only between sites belonging to different sublattices.
The resulting self-consistent Majorana hopping lattice forms a planar and bipartite graph, with an emergent square plaquette in the vicinity of the impurity site.
This allows us to compare its flux configuration with the flux pattern favored by Lieb's theorem for the corresponding quadratic Majorana hopping problem.

Figure \ref{fig: Lieb_theorem} shows a schematic illustration of the flux configurations around an impurity in the mean-field-optimized Majorana hopping lattice for both flux sectors.
The emergent square plaquette in the vicinity of the impurity always binds a flux, irrespective of the flux sector.
In the bound-flux sector, no additional fluxes appear in the surrounding hexagonal plaquettes, fully consistent with Lieb’s conjecture.
By contrast, in the zero-flux sector the deformed hexagonal plaquette also carries a $\pi$-flux.
Although this comparison is only qualitative, it suggests that the self-consistent Majorana hopping network develops a local energetic preference for the bound-flux sector. As the impurity coupling $g$ increases, this local contribution becomes increasingly important, providing a possible explanation for the reentrant transition to the bound-flux sector at large $g/J$.

\section*{Discussion}
\label{sec: discussion}
In this work, we have shown that spin-3/2 magnetic impurities provide an effective local probe of emergent flux degrees of freedom in the Kitaev spin liquid.
We demonstrated that impurity quadrupole moments encode detailed information about the surrounding flux configuration and flux sector transitions.
Using the SO(6) Majorana representation combined with a self-consistent mean-field treatment, we identified flux sector transitions through discontinuous changes in the impurity quadrupole moments. We further analyzed the effects of SIA and external magnetic fields, confirming reentrant and field-induced flux sector transitions, as well as the emergence of localized MZMs in the bound-flux sector. In addition, quadrupole correlations between impurities in a finite field exhibit exponential decay with flux-sector-dependent decay rates, reflecting the presence or absence of MZMs.

Our analysis also highlights the important role of symmetry. 
In clusters that preserve reflection symmetry, the optimized Majorana hopping lattice becomes planar and bipartite,
allowing a direct connection to Lieb’s conjecture and stabilizing impurity-bound $\pi$-fluxes.
More generally, our results establish impurity quadrupole moments as sensitive probes of emergent fluxes and fractionalized excitations in Kitaev spin liquids.

From an experimental perspective, in addition to the dilute spin-$3/2$ impurity systems discussed above \cite{Bastien2019, Lee2023, Natori2025}, another possible microscopic realization is a spin-$1$ impurity strongly and ferromagnetically coupled to a spin-$1/2$ moment in the Kitaev layer, thereby forming an effective local degree of freedom with $S_{\rm tot}=3/2$. A similar setup involving impurities with different spin quantum numbers has also been proposed in Ref. \cite{Vojta2016}.
Detecting magnetic quadrupole moments remains challenging because they do not couple linearly to either electric or magnetic fields.
Nevertheless, several theoretical studies have discussed possible probes of quadrupolar order in spin systems, including neutron \cite{Smerald2015}, Raman \cite{Michaud2011}, and resonant inelastic x-ray scattering \cite{Savary2015}, as well as nuclear magnetic resonance (NMR) experiments \cite{Sato2009, Sato2011, Shindou2013, Smerald2016}.
While these techniques have primarily been discussed in the context of bulk quadrupolar order, we have proposed that the local quadrupole moment of a spin-3/2 impurity may also be accessible through spin-polarized tunneling spectroscopy by exploiting the spectral sum rule for the on-site dynamical spin correlation.
We hope that this proposal will stimulate future experimental studies of local quadrupolar responses in candidate Kitaev materials.

\section*{Methods}
\subsection*{Derivation of mean-field Hamiltonian}
\label{sec: derivation}
To analyze the impurity problem within the Majorana representation, we perform a mean-field decoupling of the interacting terms in the impurity Hamiltonian. 
This procedure replaces quartic Majorana operators by bilinear terms multiplied by self-consistently determined mean-field parameters. 
Applying this decoupling to the impurity-host coupling $\hat{H}_\Lambda$ yields the following mean-field Hamiltonian:
\begin{align}
    \hat{H}_\Lambda\rightarrow\hat{H}_\Lambda^{\rm{MF}}=~
    & \sum_{\begin{subarray}{c} j\in\Lambda,\,k\in\partial I_j\\
    \langle jk\rangle_x
    \end{subarray}}\mathcal{U}_j^x \left[-\sum_{\rho=x,y,z}\left(\tau_j^\rho\,i\gamma_j^\rho c_{k}^{} + \mathcal{T}_{j,k}^{\rho}\, i\gamma_j^{\rho'}\gamma_j^{\rho''} - \tau_j^\rho \mathcal{T}_{j,k}^{\rho}\right)- i \gamma_j^z  c_{k}^{} + \sqrt{3} i \gamma_j^x c_{k}^{}\right]\nonumber\\
    +& \sum_{\begin{subarray}{c} j\in\Lambda,\,k\in\partial I_j\\
    \langle jk\rangle_y
    \end{subarray}}\mathcal{U}_j^y \left[-\sum_{\rho=x,y,z}\left(\tau_j^\rho\,i\gamma_j^\rho c_{k}^{} + \mathcal{T}_{j,k}^{\rho}\, i\gamma_j^{\rho'}\gamma_j^{\rho''} - \tau_j^\rho \mathcal{T}_{j,k}^{\rho}\right)- i \gamma_j^z  c_{k}^{} - \sqrt{3} i \gamma_j^x c_{k}^{}\right]\nonumber\\
    +& \sum_{\begin{subarray}{c} j\in\Lambda,\,k\in\partial I_j\\
    \langle jk\rangle_z
    \end{subarray}}\mathcal{U}_j^z \left[-\sum_{\rho=x,y,z}\left(\tau_j^\rho\,i\gamma_j^\rho c_{k}^{} + \mathcal{T}_{j,k}^{\rho}\, i\gamma_j^{\rho'}\gamma_j^{\rho''} - \tau_j^\rho \mathcal{T}_{j,k}^{\rho}\right) + 2i \gamma_j^z  c_{k}^{}\right],
    \label{eq: MF_imp_coup}
\end{align}
where $(\rho,\rho'\rho'')=(x,y,z)$ and all its cyclic permutations.
 Similarly, we obtain $\hat{H}_\kappa\rightarrow\hat{H}_\kappa^{\rm{MF}}$ around an impurity by using the following decomposition:
\begin{align}
   (0\leftarrow2):\quad\hat{\Theta}_0^x \hat{S}_1^z \hat{S}_2^y
   &=\frac18 \mathcal{U}_0^x u_{2,1}^y\,\left[i\gamma_0^{xyz}c_{2}^{} - i\gamma_0^zc_2^{} + \sqrt3 i\gamma_0^x c_2^{}\right]\nonumber\\
   &\rightarrow\frac18\mathcal{U}_0^x u_{2,1}^y\left[-\sum_{\rho=x,y,z}\left(\tau_0^\rho\,i\gamma_0^\rho c_2^{} + \mathcal{T}_{0,2}^\rho\,i\gamma_0^{\rho'}\gamma_0^{\rho''} - \tau_0^\rho\mathcal{T}_{0,2}^\rho\right)- i\gamma_0^zc_2^{} + \sqrt3 i\gamma_0^x c_2^{}\right],
\end{align}
\begin{align}
    &(1\leftarrow9):\quad\hat{S}_1^x\hat{\Theta}_0^z\hat{S}_9^y
    =\frac18(-\mathcal{U}_0^x\mathcal{U}_0^y)\times\left[ic_1^{}c_9^{}-2i^2c_1^{}\gamma_0^x\gamma_0^yc_9^{}\right]\quad{\textrm{with}}\nonumber\\
    &i^2c_1^{}\gamma_0^x\gamma_0^yc_9^{}\rightarrow\left[\tau_0^z\,ic_1^{}c_9^{} +t_{1,9}\,i\gamma_0^x\gamma_0^y - \tau_0^zt_{1,9}
    -(\mathcal{T}_{0,1}^x\,i\gamma_0^yc_9^{} + \mathcal{T}_{0,9}^y\,i\gamma_0^xc_1^{} - \mathcal{T}_{0,1}^x\mathcal{T}_{0,9}^y)\right.\\
    &\qquad\qquad\qquad\qquad\qquad\qquad\qquad\qquad\qquad\qquad\qquad\left.+\mathcal{T}_{0,9}^x\,i\gamma_0^yc_1^{}
    +\mathcal{T}_{0,1}^y\,i\gamma_0^xc_9^{}-\mathcal{T}_{0,9}^x\mathcal{T}_{0,1}^y
    \right]\nonumber
\end{align}
Here, we use \textit{relative} site labels around an impurity, as illustrated in Fig.~\ref{fig: internal_flux}, and ($j\leftarrow k$) denotes Majorana hopping from site $j$ to site $k$ driven by the $\kappa$ term.
To evaluate the second-type coupling terms, we express the impurity operators projected onto the physical Hilbert space, $\hat{\Delta}\hat{\Theta}^\mu$, in terms of the SO(6) Majorana fermions $\beta^\mu$ and $\gamma^\mu$ as introduced in  Eq. \eqref{eq: SO(6)_Majorana}:
\begin{align}
\begin{cases}
    \hat{\Delta}\hat{\Theta}^x
    &= \frac{i}{2}\beta^y\beta^z - \frac{1}{2}\beta^y\beta^z\gamma^x\gamma^y + \frac{\sqrt3}{2}\beta^y\beta^z\gamma^y\gamma^z,\\
    \hat{\Delta}\hat{\Theta}^y
    &= \frac{i}{2}\beta^z\beta^x - \frac{1}{2}\beta^z\beta^x\gamma^x\gamma^y - \frac{\sqrt3}{2}\beta^z\beta^x\gamma^y\gamma^z,\\
    \hat{\Delta}\hat{\Theta}^z
    &= \frac{i}{2}\beta^x\beta^y + \beta^x\beta^y\gamma^x\gamma^y.
\end{cases}
\end{align}
 The number of independent mean-field parameters depends on the number of impurities introduced in a given cluster.
In zero magnetic field, we treat 12 independent mean-field parameters per impurity, namely $\tau_j^\mu$ and $\mathcal{T}_{j,k}^\mu$ for $\mu=x,y,z$, where $k$ runs over the three NN sites of an impurity site $j$.
When a magnetic field is applied, the number of mean-fields increases to 33 per impurity due to the inclusion of NNN hoppings, such as $\mathcal{T}_{j,k}^\mu$ with $\langle\!\langle jk\rangle\!\rangle$ and $t_{k,l}$, in addition to the mean-fields considered in the zero-field case.
Note that some mean-fields are symmetry-forbidden for acquiring finite values: $\tau^x$ is prohibited by the reflection symmetry and $\tau^y$ by TRS.
We observe these properties numerically, as detailed in the Results section.

We treat all gauge fields, including $\mathcal{U}_j^\mu$, as static variables that can only take the values $\pm 1$. The resulting Hamiltonian then involves only the matter Majorana fermions. In this approach, one can evaluate the ground-state energies for all possible gauge configurations and select the one that yields the lowest energy to discuss physical observables in the ground state. 
Note that we do not omit constant terms when evaluating the ground-state energy for different flux configurations.
Mean-field parameters are optimized self-consistently via the Steffensen method.
For each parameter set, the optimization starts from random initial mean-field parameters, and the continuity of the converged solution under parameter sweeps is used to exclude metastable solutions.
Each optimization process ends when the norm of the mean-field difference is less than $10^{-9}$.

\subsection*{Benchmark}
\begin{figure}[b]
    \centering
    \includegraphics[width=0.95\linewidth]{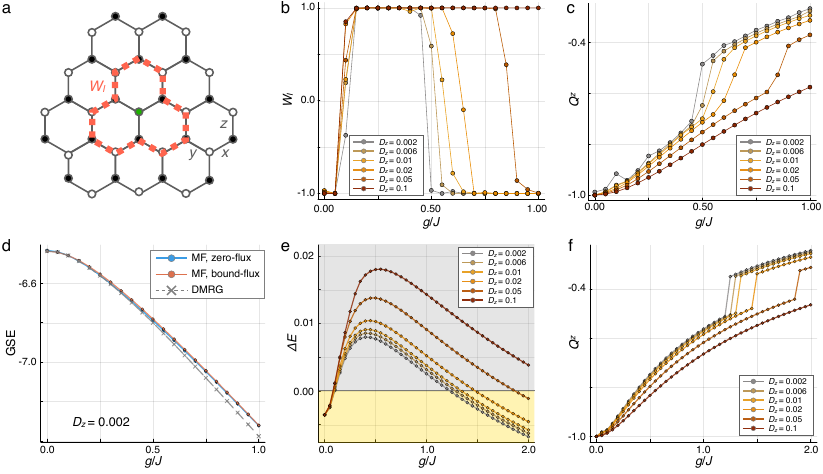}
    \caption{
    \textbf{Benchmark of the mean-field approximation for a 37-spin-site cluster with open-boundary conditions.}
    \textbf{a} Cluster geometry used in the benchmark. The impurity site is located at the center of the cluster (green).
    \textbf{b} Spin-DMRG results for the triple-plaquette operator, characterized by $W_I$, as a function of $g/J$ for several values of $D_z$.
    \textbf{c} Impurity quadrupole moment $Q^z$ obtained by spin-DMRG for the same cluster.
    \textbf{d} Ground-state energies (GSEs) obtained by the Majorana mean-field (MF) calculation in the zero- and bound-flux sectors, together with the spin-DMRG result, for $D_z=0.002$.
    \textbf{e} Energy difference $\Delta E = E_{\rm bound}-E_{\rm zero}$ between the bound- and zero-flux sectors obtained by the Majorana mean-field calculation.
    \textbf{f} Impurity quadrupole moment $Q^z$ obtained by the Majorana mean-field calculation.}
    \label{fig:benchmark_37site}
\end{figure}

We benchmark our mean-field approximation against spin-DMRG calculations for a 37-spin-site cluster with open-boundary conditions, as shown in Fig.~\ref{fig:benchmark_37site}a. The impurity site is located at the center of the cluster, highlighted in green. No magnetic field is applied.
The spin-DMRG calculations were performed using ITensors.jl \cite{ITensor} with a maximum bond dimension of $2^{12}$, a truncation cutoff of $10^{-9}$, and a convergence criterion of $10^{-7}$ for the energy difference between successive sweeps.

We first summarize the spin-DMRG results. Although $W_I=\pm1$ in principle, finite numerical accuracy yields nonquantized values around transition points. The system exhibits a reentrant transition into the bound-flux sector around $g/J\sim1$ for small $D_z$, with $g_2\propto D_z$ as shown in Fig. \ref{fig:benchmark_37site}b. Correspondingly, $Q^z$ exhibits a discontinuous jump across the transition (see Fig. \ref{fig:benchmark_37site}c). 

We next compare the spin-DMRG and mean-field results. Within the mean-field framework, the ground-state energy can be evaluated separately for each flux sector, assuming even matter-fermion parity in both sectors. Figure \ref{fig:benchmark_37site}d shows the ground-state energies for $D_z=0.002$ as a function of $g/J$. Although the agreement is good at small $g/J$, the deviation from the DMRG result increases with increasing $g/J$.
We also confirm that, in the limit of $D_z\rightarrow\infty$, the zero-flux mean-field result converges to the DMRG result.

The energy difference between the two flux sectors, $\Delta E$, is shown in Fig. \ref{fig:benchmark_37site}e. The mean-field calculation reproduces the qualitative reentrant transition observed in the DMRG results, although the second transition point, $g_2(D_z)$, is quantitatively overestimated. Figure \ref{fig:benchmark_37site}f shows the corresponding mean-field results for $Q^z$, whose magnitude is in good quantitative agreement with the DMRG results. Overall, the mean-field approximation captures the qualitative physics and provides a good quantitative estimate of the quadrupole moment, but is less accurate in predicting the transition points.

\section*{Data availability}
The datasets generated during and/or analyzed during the current study are available from the corresponding author upon reasonable request.

\section*{Code availability}
The codes used during the current study are available from the corresponding author upon reasonable request.

\section*{Acknowledgment}
We thank J. Knolle for fruitful discussions.
M.O.T. is supported by a Japan Society for the Promotion of Science (JSPS) Fellowship for Young Scientists. Support for W.-H.K. was provided by the Office of the Vice Chancellor for Research and Graduate Education at the University of Wisconsin-Madison with funding from the Wisconsin Alumni Research Foundation. N.B.P. were supported
by the U.S. Department of Energy, Office of Science, Basic Energy Sciences under Award No. DE-SC0018056.  
This work is supported by JSPS KAKENHI No. JP25KJ0409, No. JP25H00609, and No. JP25K22011.
Some of the numerical calculations have been carried out on the HOKUSAI at RIKEN.

\section*{Author Contributions}
M.O.T. and N.B.P. devised the project.
M.O.T. carried out all calculations in this study.
M.O.T., W.-H.K., S.F., and N.B.P. contributed to the interpretation of the results and the writing of the paper.

\section*{Competing interests}
The authors declare no competing interests.

\bibliography{paper}
\end{document}